\documentclass[conference]{IEEEtran}

\usepackage{cleveref}
\usepackage{amsmath}
\usepackage{enumerate}
\usepackage{amsfonts}
\usepackage{mathtools}

\newtheorem{example}{Example}

\newtheorem{proposition}{Proposition}
\newtheorem{corollary}{Corollary}
\newtheorem{definition}{Definition}
\newtheorem{lemma}{Lemma}
\newtheorem{theorem}{Theorem}

\newcommand{\F}{\mathbb{F}}
\newcommand{\T}{\ensuremath{\mathcal{T}}}
\newcommand{\PowSet}{\ensuremath{\mathcal{P}}}
\DeclareMathOperator{\supp}{supp}
\DeclareMathOperator{\rep}{Rep}
\DeclareMathOperator{\diag}{Diag}
\DeclareMathOperator{\rank}{rk}

\begin{document}
	%
	% paper title
	% Titles are generally capitalized except for words such as a, an, and, as,
	% at, but, by, for, in, nor, of, on, or, the, to and up, which are usually
	% not capitalized unless they are the first or last word of the title.
	% Linebreaks \\ can be used within to get better formatting as desired.
	% Do not put math or special symbols in the title.
	\title{Private Information Retrieval Schemes for Coded Data  % from Colluding Servers 
	with Arbitrary Collusion Patterns}

	% author names and affiliations
	% use a multiple column layout for up to three different
	% affiliations
%	\author{\IEEEauthorblockN{O.~W.~Gnilke, D.~Karpuk, R.~Freij-Hollanti, C.~Hollanti
%		}
%		\IEEEauthorblockA{
%			%Algebra, Number Theory, and Applications research group\\
%			Department of Mathematics and Systems Analysis\\
%			Aalto University, School of Science, Finland\\
%			Emails: \{oliver.gnilke, david.karpuk, ragnar.freij, camilla.hollanti\}@aalto.fi}
%		\and
%		\IEEEauthorblockN{R.~Tajeddine, S.~El Rouayheb }
%		\IEEEauthorblockA{ECE Department, IIT, Chicago\\
%			Email: : rtajeddi@hawk.iit.edu, salim@iit.edu}
%	}
	
	% conference papers do not typically use \thanks and this command
	% is locked out in conference mode. If really needed, such as for
	% the acknowledgment of grants, issue a \IEEEoverridecommandlockouts
	% after \documentclass
	
	% for over three affiliations, or if they all won't fit within the width
	% of the page, use this alternative format:
	% 
	\author{\IEEEauthorblockN{Razane Tajeddine\IEEEauthorrefmark{2},
	Oliver W.~Gnilke\IEEEauthorrefmark{1},
	David Karpuk\IEEEauthorrefmark{1}, 
	Ragnar Freij-Hollanti\IEEEauthorrefmark{1},\\
	Camilla Hollanti\IEEEauthorrefmark{1} and
	Salim El Rouayheb\IEEEauthorrefmark{2}}
	\IEEEauthorblockA{\IEEEauthorrefmark{2}ECE Department, IIT, Chicago\\
		Email:  rtajeddi@hawk.iit.edu, salim@iit.edu}
	\IEEEauthorblockA{\IEEEauthorrefmark{1}%Algebra, Number Theory, and Applications research group\\
		Department of Mathematics and Systems Analysis\\
		Aalto University, School of Science, Finland\\
		Email: \{oliver.gnilke, david.karpuk, ragnar.freij, camilla.hollanti\}@aalto.fi}}
	%\IEEEauthorblockA{\IEEEauthorrefmark{3}Starfleet Academy, San Francisco, California 96678-2391\\
	%Telephone: (800) 555--1212, Fax: (888) 555--1212}
	%\IEEEauthorblockA{\IEEEauthorrefmark{4}Tyrell Inc., 123 Replicant Street, Los Angeles, California 90210--4321}

	% use for special paper notices
	%\IEEEspecialpapernotice{(Invited Paper)}

	% make the title area
	\maketitle
	
	% As a general rule, do not put math, special symbols or citations
	% in the abstract
	\begin{abstract}
	In Private Information Retrieval (PIR), one wants to download a file from a database without revealing to the database which file is being downloaded.  Much attention has been paid to the case of the database being encoded across several servers, subsets of which can collude to attempt to deduce the requested file.  With the goal of studying the achievable PIR rates in realistic scenarios, we generalize results for coded data from the case of all subsets of servers of size $t$ colluding, to arbitrary subsets of the servers.  We investigate the effectiveness of previous strategies in this new scenario, and present new results in the case where the servers are partitioned into disjoint colluding groups.
	\end{abstract}
	
	% no keywords

	% For peer review papers, you can put extra information on the cover
	% page as needed:
	% \ifCLASSOPTIONpeerreview
	% \begin{center} \bfseries EDICS Category: 3-BBND \end{center}
	% \fi
	%
	% For peerreview papers, this IEEEtran command inserts a page break and
	% creates the second title. It will be ignored for other modes.
	\IEEEpeerreviewmaketitle

		% !TEX root = Salim_ISIT2017_PIR_collusion_patterns.tex
	
	\section{Introduction}
	% no \IEEEPARstart
		In Private Information Retrieval (PIR), a user is interested in obtaining the data in an entry of an online database, while keeping private which entry she is interested in. A straightforward solution that achieves perfect privacy consists of the user downloading the whole database. %However, this solution is impractical especially with the large size of databases in most applications.  
		In their  seminal paper \cite{PIR1995, chor1998private}, Chor \emph{et al.} introduced the concept of PIR and devised PIR schemes with substantially lower communication cost (sublinear in the size of the database) that achieve perfect privacy in an information theoretic sense. The key idea was to replicate the database on multiple servers and assume that the servers do not collude. Clearly, if the data is accessible only from a single server, there is no better scheme than downloading the whole database for achieving (information theoretic) perfect privacy.
		%\footnote{mention computation PIR \cite{CompPIR}.  say they are not efficient since they rely on homomorphic encryptio }.		

%\noindent
{\em Replication-based PIR:} Since its introduction, PIR has received ample attention, and the literature has focused mainly on minimizing the (total) communication cost in the replicated databases model. In recent breakthrough results, PIR schemes with subpolynomial communication cost have been constructed for multiple servers \cite{yekhanin2008towards,efremenko20123, beimel2002breaking} and later for two servers \cite{dvir20142}. 
%A parallel line of work on PIR focused on achieving privacy in a computational sense based on the assumed hardness of certain problems. Computational PIR schemes that    require a single server were first constructed in  \cite{kushilevitz1997replication}. The most efficient scheme requires a  polylogarithmic communication cost \cite{cPIRPoly}. However, these schemes suffer from   very high computational complexity.     In this paper, we are only concerned with informational theoretic privacy.

%\noindent
{\em Coding-based PIR:} Furthermore, there has been a growing interest in studying PIR schemes for coded data, due to the savings in the storage overhead compared to mere replication. Shah \emph{et al.} showed in \cite{shah2014one} that only one extra bit download is needed to achieve privacy when the data is coded on an exponentially large number of servers. Blackburn \emph{et al.} achieved the same low download complexity with a linear number of servers~\cite{blackburn2016small}. Fazeli \emph{et al.} \cite{fazeli2015pir} focused on the design of codes that minimize the storage overhead of PIR and provided a method to transform any linear replication-based PIR scheme into a scheme for coded data. %Tajeddine et al. provided explicit PIR schemes for privately retrieving MDS coded files from a distributed storage systems in \cite{tajeddine2016private,Extended}. These schemes were subsequently unified as special cases of a more extensive family of PIR schemes in~\cite{FGHK16}. 
In addition to various constructions, the fundamental limits on the download cost of PIR schemes have been characterized for replicated data \cite{sun_jafar_1} and coded data \cite{bananaman}.
%\cite{chan2014private,augot2014storage,fazeli2015pir}. However,  these works  only focus on the communication cost and   are either non-constructive \cite{chan2014private} or the PIR schemes do not work with existing codes for DSS. For instance, the results in \cite{shah2014one} require the number of servers $n$ to be exponentially large in the number of files, and the constructions in \cite{augot2014storage,fazeli2015pir} require encoding together    data across different files, which may not be always possible in DSSs. Typically in DSSs,  only data chunks within the same file are allowed to be encoded together. The PI has started addressing these problems in \cite{tajeddine2016private,Extended}. 

%\noindent
{\em Collusion:}  In replication-based PIR, the model that has received most attention is non-colluding servers. Here the assumption is that the nodes (\emph{e.g.}, servers)
 do not communicate and are therefore only aware of the query they received from the user. Distributed storage systems on the other hand depend upon communication between nodes to facilitate repair functions. This motivated investigating PIR in the context of limited collusion, \emph{i.e.}, a subset of servers can share their knowledge in an attempt to deduce the user's request. We refer to this as \emph{$t$-PIR}, meaning that any subset of up to $t$ servers is allowed to collude. The capacity for $t$-PIR has been established in \cite{sun2016capacity}. For coded $t$-PIR  the capacity is yet unknown, but \cite{razan_salim} described a first scheme for $t$-PIR on coded databases employing a maximum distance separable (MDS) code. In \cite{FGHK16} this scheme was extended to a wider set of parameters and an algebraic framework was established, also resulting in a conjecture for the coded $t$-PIR capacity.

%\noindent
{\em Contributions:} 
To the best of our knowledge we present the first work on PIR schemes on coded data for arbitrary collusion patterns. We describe a strategy that employs a $t$-PIR scheme for a minimal $t$, for any given collusion pattern. We continue to examine some special cases of collusion patterns that allow for a more tailored approach that increases the rate significantly. Most notable is the case of the collusion pattern not being connected, that is, being contained in a partition of the set of nodes. In this case an approach that resembles schemes for collusion-free PIR can be utilized and almost equally high rates achieved. This case is of special interest since it can represent server groups that are geographically or otherwise separated.

		\section{Preliminary Definitions}

Let us describe the distributed storage systems we consider; this setup follows that of \cite{razan_salim,FGHK16,bananaman}.  To provide clear and concise notation, we have consistently used superscripts to refer to files, subscripts to refer to servers, and parenthetical indices for entries of a vector.  So, for example, the query $q^i_j$ is sent to the $j^{th}$ server when downloading the $i^{th}$ file, and $y_j(a)$ is the $a^{th}$ entry of the vector stored on server $j$.

Suppose we have files $x^1,\ldots,x^m\in \mathbb{F}_q^k$.  The considered data storage scheme proceeds by arranging the files into an $m \times k$ matrix
\begin{equation} \
X=
\left[\begin{smallmatrix}
x^1 \\ \vdots \\ x^m
\end{smallmatrix}
\right]=
\left[\begin{smallmatrix}
	x^1(1) & \cdots & x^1(k) \\
		\vdots & \ddots	& \vdots \\
	x^m(1) & \cdots & x^m(k)\\
\end{smallmatrix}\right].
\end{equation}
Each file $x^i$ is encoded using a linear $[n,k,d]$-code $C$ with length $n$, dimension $k$, and minimum distance $d$, and having generator matrix $G_C$, into an encoded file $y^i = x^iG_C$.  In matrix form, we encode the matrix $X$ into a matrix $Y$ by right-multiplying by $G_C$:
\begin{equation} Y= X G_C = \left[
\begin{smallmatrix}
y^1 \\ \vdots \\ y^m
\end{smallmatrix}\right]
= 
\left[\begin{smallmatrix}
y_1 & \cdots & y_n
\end{smallmatrix}\right]\,.
\end{equation}
The $j^{th}$ column $y_j\in\mathbb{F}_q^m$ of the matrix $Y$ is stored by the $j^{th}$ server.  Such a storage system allows $d-1$ servers to fail while still allowing users to successfully access any of the files $x^i$.  If $C$ is an MDS code, the resulting distributed storage system is maximally robust against server failures.

	\begin{definition}\label{PIR_def}
Suppose we have a distributed storage system as above, where $m$ files are stored across $n$ servers.  A \emph{PIR scheme} for such a storage system consists of:
%\begin{small}
\begin{itemize}
\item[1.] For each index $i\in[m]$, a probability space $(\mathcal{Q}^i,\mu^i)$ of \emph{queries}.  When the user wishes to download $x^i$, a query $q^i \in \mathcal{Q}^i$ is selected randomly according to the probability measure $\mu^i$.  Each $q^{i}$ is itself a tuple $q^{i} = (q^{i}_1,\ldots,q^{i}_n)$, where $q^{i}_j$ will be sent to the $j^{th}$ server. Here, $q^{i}_j$ is a function that takes the content of server $j$ as input, and returns one symbol in the alphabet of the code.
\item[2.] Responses $r^{i}_j = q^{i}_j(y_j)$ which the servers compute and transmit to the user.
\item[3.] A reconstruction function which takes as inputs the $r^{i}_j$ and returns some $c$ coordinates of the $i^{th}$ file.
\end{itemize}
%\end{small}
\end{definition}

	We call a set $T \subseteq [n]$ a \emph{collusion set} if it is possible for the servers indexed by $T$ to share their requests in an attempt to deduce the index of the requested file. We denote the collection of all colluding sets $\T$, and call it a \emph{collusion pattern}. Clearly, any subset of a colluding set is again colluding, and so $\T$ is closed under inclusion, or in combinatorial terms an abstract simplicial complex. We can therefore describe $\T$ by its set of maximal elements, and we write $\T=\langle T_1,\cdots , T_r\rangle$ if $T_1,\cdots , T_r$ are the maximal colluding sets.

\begin{definition}
A collusion pattern $\T$ is \emph{disconnected} if there exist non-empty disjoint sets $T_1$ and $T_2$ such that $\T\subseteq\langle T_1,T_2\rangle$.
\end{definition}

\begin{definition}\label{collusion}
A PIR scheme \emph{protects against the colluding set $T=\{j_1,\ldots,j_t\}\subseteq[n]$} if we have
\begin{equation}\label{mi}
I(q^{i}_{j_1},\ldots,q^{i}_{j_t};i) = 0
\end{equation}
where $I(\cdot\ ;\cdot)$ denotes the mutual information of two random variables.  In other words, there exists a probability distribution $(\mathcal{Q}_T,\mu_T)$ such that, for all $i\in [m]$, the projection of $(\mathcal{Q}^i,\mu^i)$ to the coordinates in $T$ is $(\mathcal{Q}_T,\mu_T)$. Hence, the servers in $T$ will not learn anything about the index $i$ of the file that is being requested. A PIR scheme that protects against all sets in a collusion pattern $\T$ is said to be $\T$-secure.
\end{definition}
	
	Clearly, if $S\subseteq T$, then any PIR scheme that protects against $T$ also protects against $S$. Moreover, if $\mathcal{S}\subseteq \T$ are two collusion patterns, then any $\T$-secure PIR scheme is also $\mathcal{S}$-secure.
	
	For the rest of this paper we will exclusively consider linear schemes that use uniform distributions, as in the following fundamental example.
	\begin{example}
		Let $n=2$ servers each store a copy of a database consisting of $m$ files $x^\ell \in \F_q$. To retrieve the $i^{th}$ file the user chooses uniformly at random an element $u$ in $\F_q^m$ and constructs the queries as $q^i=(q^i_1,q^i_2)=(u,u+e_i)$. The space of all queries therefore is given by $\mathcal{Q}^i=\{(x,y) : y-x = e_i\}$ and $\mu^i=\frac{1}{q^n}$ is the uniform probability measure. The responses $r^i_j:=\langle x,q^i_j \rangle $ are calculated as the inner product of the database with the requests and reconstruction is achieved by subtraction of the responses, $x^i=r^i_2-r^i_1$.
		
		This scheme is $\{\{1\},\{2\}\}$-secure, as both projections of any query space $\mathcal{Q}^i$ onto a coordinate are identical to the complete ambient space $\F_q^n$ with uniform measure. It does not, however, protect against collusion in $\{1,2\}$, as the two servers between them can observe the index $i$ from the difference between their query vectors.
	\end{example}

		\section{A generic PIR scheme for colluding sets}
	
	In this section we briefly summarize the methods of \cite{FGHK16}, which constructed explicit PIR schemes for the case where $\T =\{ T : |T|\leq t \}\stackrel{\rm{def}}{=}\binom{n}{\leq t}$, meaning that all of the subsets of $[n]$ of size $t$ are colluding.  A scheme that is a $\binom{n}{\leq t}$-secure will be called a \emph{$t$-PIR scheme}. The crucial ingredients to the PIR schemes discussed in \cite{FGHK16} are the following: the storage code $C\subseteq \F^n$, another linear code $D$ of the same length $n$ as $C$, and the \emph{star product}
	\[
	C\star D = \text{span}\{[c(1)d(1),\ldots,c(n)d(n)]\in \F^n: c\in C, d\in D\}
	\]
of the linear codes $C$ and $D$, which is again a linear code of length $n$.  The main theorem of \cite{FGHK16} is the following:

\begin{theorem}[\cite{FGHK16}]
Given a storage code $C$ and a linear code $D$ as above, there exists a linear PIR scheme for the distributed storage system $Y = XG_C$ with rate $(d_{C\star D}-1)/n$ which protects against all colluding sets of size $d_{D^\perp}-1$.
\end{theorem}

		The schemes in \cite{razan_salim} and \cite{FGHK16} allow to retrieve linear combinations of files, or even parts of different files. For simplicity we will henceforth restrict to the case where a single file is retrieved.
		
		To privately retrieve a file $x^i$, for every file $x^\ell$ in the database a codeword $d^\ell$ is chosen uniformly at random from the code $D$. A vector $e \notin D$ is then added to $d^i$.  The query $q_j^i\in \F^m$ sent to the $j$:th server is then
		\[
		q^i_j = [d^1(j),\ldots,d^i(j)+e(j),\ldots,d^m(j)]
		\]
and the servers respond with
		\[
		\begin{split}
		[r^i_1,\ldots,r^i_n] &= [\langle q^i_1,y_1 \rangle,\ldots,\langle q^i_n,y_n\rangle] \in C\star D + C\star e.
		\end{split}
		\]
The support of $e$ is then chosen so that decoding the vector $[r^i_1,\ldots,r^i_n]$ to its closest neighbor in $C\star D$ reveals $d_{C\star D}-1$ coordinates of $y^i$, coming from the $C\star e$ summand in the above expression.
	We will refer to the above as a \emph{$(D,e)$-retrieval scheme}, where $D$ and $e$ are as above.

	\begin{example}\label{ex:fullcoll}
	Suppose that $1\leq t\leq n -k$.  By choosing $C$ and $D$ to both be generalised Reed-Solomon (GRS) codes with the same evaluation vector, the $(D,e)$-retrieval scheme of \cite{FGHK16} can achieve a rate of $\frac{n-(k+t-1)}{n}$ while protecting against all colluding sets of size $t$.  See \cite{FGHK16} for more details.
	\end{example}
	
	\begin{example}\label{ex:rep}
	Suppose we let $D = \rep(n)_q$ be the repetition code of length $n$ over $\F$.  A corresponding $(D,e)$-retrieval scheme then admits a particularly simple description, first described in \cite{razan_salim}.  Specifically, pick a vector $u\in \F^{m}$ uniformly at random.  To server $j = 1,\ldots,n$, send the query
	\[
	q_j^i = \left\{
	\begin{array}{cl}
	u + e_i & \text{if $j\in \supp(e)$} \\
	u & \text{otherwise}
	\end{array}
	\right.
	\]
	where $e_i$ is the $i$:th standard basis vector.  The vector $r$ consisting of all the responses satisfies
	\[
	r = [r^i_1,\ldots,r^i_n] \in C\star \rep(n) + C\star e = C + C\star e
	\]
	because $C\star \rep(n) = C$.  Thus provided $|\supp(e)|<d_C$, the $(D,e)$-retrieval scheme has rate $\frac{n-k}{n}$. However, it only protects against colluding sets of size $t = 1$, that is, no non-trivial collusion.  This strategy works for any linear code $C$.
	\end{example}

Hence, in the case where all $t$-sets are colluding, the $(\rep, e)$-scheme from Example~\ref{ex:rep} is only secure in the trivial case of $t=1$. However, we will see in Corollary~\ref{thm:disconn} that for partial collusion, the repetition code plays a much more prominent role.	

	\section{Partial Collusion}
%	We call a set $T \subseteq [n]$ a collusion set if it is possible for the servers indexed by $T$ to share their requests in an attempt to deduce the requested file. We denote the collection of all colluding sets $\T$. Clearly, any subset of a colluding set is again colluding, and so $\T$ is closed under inclusion, or in combinatorial terms an abstract simplicial complex. We can therefore describe $\T$ by its set of maximal elements.
%	
%	The scheme described in \cite{FGHK16} chooses for every file in the database a codeword $d^\ell$ from a linear code $D$. For a file $w$ that the user wants to retrieve, a vector $e \notin D$ is added to $d^w$. We will call this scheme the \emph{$(D,e)$-retrieval scheme}, where $D$ is a linear code of length $n$, and $e$ is a vector of length $n$. 

In general, the security of the $(D,e)$-retrieval scheme depends upon the following observation. 
	
	\begin{proposition}\label{SecuritytPIR}
		The PIR scheme in \cite{FGHK16} is secure against collusion of the set $T$, if the projection of the vector $e$ lies in the projection of D, \emph{i.e.}, $e_T \in D|_T$.
		\begin{IEEEproof}
			The query for any file that is not requested is a uniform vector $d\in D$. The colluding servers in the set $T$ see a projection of these codewords onto their coordinates, hence a uniform element in $D|_T$. The security relies on indistinguishability of the query for non-requested and requested files. It follows that the PIR scheme is secure if $d+e_T \in D|_T$. As $D|_T$ is a linear code, this is equivalent to the condition $e_T \in D|_T$. 
		\end{IEEEproof}
	\end{proposition} 
	\begin{corollary}\label{thm:disconn}
		Let the storage code $C$ be MDS. There is a vector $e$ such that the $(\rep,e)$-retrieval scheme is $\T$-secure and has positive retrieval rate if and only if $\T$ is disconnected.
	\begin{IEEEproof}
		 Assume that the $(\rep,e)$-retrieval scheme has positive retrieval rate. In particular, we must have $e\notin \rep$, as otherwise the response vector would be in $C\star\rep=C$, and would be annihilated when decoding with $C^\perp$. Let $T_1=\{i\in [n]:e(i)=e(1)\}$. By construction $1\in T_1$, so $T_1$ is non-empty. On the other hand, as $e$ is not in the repetition code, the set $T_2=T_1^c=\{i\in [n]:e(i)\neq e(1)\}$ is also non-empty. If the $(\rep,e)$-scheme is $\T$-secure, then $e_T$ is a repetition vector for all $T\in\T$, so in particular $T$ can not intersect both $T_1$ and $T_2$. Thus $\T$ is a disconnected collusion pattern.
		 
On the other hand, let $\T\subseteq\langle T_1, T_2\rangle$ be a disconnected collusion pattern, where $T_1$ and $T_2$ are non-empty sets with $T_1\cap T_2=\emptyset$. Let \[e(i)=\left\{\begin{array}{ll}1& \text{if } i\in T_1\\ 0& \text{if } i\in T_0.
\end{array}\right.\] Then we retrieve a vector from $(C\star e)C^\perp$. Since $C\star e\not\subseteq C$, this code has positive rank, so there is non-trivial information downloaded. 
	\end{IEEEproof}
	\end{corollary}

%	In \cite{FGHK16} the special case $\T =\{ T : |T|\leq t \}\stackrel{\rm{def}}{=}\binom{n}{\leq t}$ is discussed. A $(D,e)$-retrieval scheme that protects against  $\binom{n}{\leq t}$ is called a $t$-PIR scheme.
	\begin{corollary}
		If the $(D,e)$-retrieval scheme is a $t$-PIR scheme, then $\dim(D)$ is at least $t$.
	\begin{IEEEproof}
		 Assume $\dim(D)<t$ and let $I$ be an information set, \emph{i.e.}, $|I|=\dim(D)$ and $D|_I$ is of full rank. The coordinates $e_I$ uniquely determine a codeword $d$ from $D$, if we consider the collusion sets $I \cup {j}$ for all $j \in [n]$ we see that for at least one such set $e(j) \neq d(j)$, because otherwise we would have $e \in D$.
	\end{IEEEproof}
	\end{corollary}

	\section{Solving partial collusion using $t$-PIR}\label{sec:partial}
	For any collusion set $\T$, we can let $t$ be the largest size of a colluding set in $\T$. Then, $\T \subseteq \binom{n}{\leq t}$, so we get a $\T$-secure retrieval scheme from the $\binom{n}{\leq t}$-secure scheme in the previous section.
	
	In \cite{FGHK16, razan_salim} it was shown that $t$-PIR schemes achieve rates of $\frac{n-k-t+1}{n}$ if the storage code $C$ is of dimension $k$ and properly chosen. We are now interested in variations of $\T$ for which a different strategy leads to capacity gains.
	
	We begin with an instructive example, adapted from~\cite{FGHK16}:
	\begin{example}\label{Example1}
	Consider the $[5,2]$-storage code where each file $x^i$ is divided into two blocks $x^i(1)$ and $x^i(2)$, and distributed onto the five servers via right-multiplication by the generator matrix: 
	\[G_C=\begin{bmatrix} 1 & 0 & 4 & 3 & 2 \\
	0 & 1 & 2 & 3& 4\end{bmatrix}\]
%\begin{small}
% \begin{equation}\label{eq:Storage} \left\{\begin{array}{ll}
% x^i(1) & \mbox{on server } 1 \\
% x^i(2) & \mbox{on server } 2 \\
%  4x^i(1)+2x^i(2) & \mbox{on server } 3 \\
%   3x^i(1)+3x^i(2) & \mbox{on server } 4 \\
%   2x^i(1)+4x^i(2) & \mbox{on server } 5.
%\end{array}
%	\right.
%	 \end{equation}	 
%	 \end{small}
This is a GRS storage code of rank $2$ over any field of characteristic $\geq 5$. If we let $D$ be the matrix  \[ G_D:=\begin{bmatrix}
		1 & 1 & 1 & 1 & 1\\
		0 & 1 & 2 & 3 & 4
	\end{bmatrix}.
	 \] and $e=(1,1,0,0,0)$, then the $(D,e)$-retrieval scheme has the following explicit expression:
	 
	 %	 For each file index $\ell\in[m]$, we sample uniformly at random from $D$ by multiplying $G_D$ on the left by a uniform random vector $z^\ell = (z^\ell(1),z^\ell(2))\in \F_5^2$, so that $d^\ell = z^\ell G_D$ and $d_j = (d^1(j),\ldots,d^m(j))$ for $j \in [m]$.  We let $z_1 = (z^\ell(1),\ldots,z^m(1))$ and $z_2 = (z^1(2),\ldots,z^m(2))$ which are independent and uniformly distributed over $\F_5^m$.
	 
The queries $q^{i}$ sent to the servers are the following vectors in $\F_5^m$: 
\begin{small}
\begin{equation}
\begin{aligned}
q^i_1 &= [z^1(1)&,\ldots &, z^m(1)] & + & e_i \\
q^i_2 &= [z^1(1) + z^1(2)&,\ldots &, z^m(1) + z^m(2)] & + & e_i \\
q^i_3 &= [z^1(1) + 2z^1(2)&,\ldots &, z^m(1) + 2z^m(2)] \\
q^i_4 &= [z^1(1) + 3z^1(2)&,\ldots &, z^m(1) + 3z^m(2)] \\
q^i_5 &= [z^1(1) + 4z^1(2)&,\ldots &, z^m(1) + 4z^m(2)]
\end{aligned}
\end{equation}
\end{small}
	  %\begin{equation}\label{eq:Queries} \left\{\begin{array}{ll}
%\sum d_j(1)\mathbf{e}_j + \mathbf{e}_i & \mbox{to server } 1 \\
%\sum (d_j(1)+d_j(2))\mathbf{e}_j + \mathbf{e}_i & \mbox{to server } 2 \\
%  \sum (d_j(1)+2d_j(2))\mathbf{e}_j & \mbox{to server } 3 \\
% \sum (d_j(1)+3d_j(2))\mathbf{e}_j & \mbox{to server } 4 \\
%  \sum (d_j(1)+4d_j(2))\mathbf{e}_j & \mbox{to server } 5,
%\end{array}
%	\right.
%	 \end{equation}	 
	 %where all sums are taken over $j\in[m]$. 
	 where $e_i$ is the $i^{th}$ standard basis vector, and $z^i(1)$ and $z^i(2)$ are random elements in $\F_5$.  
	 The servers now respond by projecting their stored data onto the query vector. %whence we obtain a response vector 
%	 \begin{equation}\label{eq:Response} r^i = \left[\begin{smallmatrix}
%\sum_{\ell = 1}^m d^\ell(1)x^\ell(1) &+ x^i(1)\\
%\sum_{\ell = 1}^m (d^\ell(1)+d^\ell(2))x^\ell(2) &+ x^i(2) \\
%  \sum_{\ell = 1}^m (d^\ell(1)+2d^\ell(2)) (4x^\ell(1)+2x^\ell(2))& \\
% \sum_{\ell = 1}^m (d^\ell(1)+3d^\ell(2))(3x^\ell(1)+2x^\ell(2)) &\\
%  \sum_{\ell = 1}^m (d^\ell(1)+4d^\ell(2))(2x^\ell(1)+4x^\ell(2))&
%\end{smallmatrix}\right].
%	 \end{equation}	 
	 It is shown in~\cite{FGHK16} how these responses can be used to decode both blocks from the desired file $x_i$. Observe that for each pair of servers, the corresponding joint distribution of  queries is the uniform distribution over $(\F_5^m)^2$. However, as a bonus, the same is true for the joint distribution of $(q_3,q_4,q_5)$. Hence, the described scheme is $\T$-secure, where $\T=\binom{5}{\leq 2}\cup\{\{3,4,5\}\}$.
	\end{example}
	
		More generally, consider the $(D,e)$-retrieval scheme used to protect against collusion of all $t$-sets, as in Example~\ref{ex:fullcoll} and call the support of the vector $e$ the \emph{information set} $\supp(e)=I$. It is obvious that any colluding set $S$ that is disjoint from $I$ can not recover any information, as $q^i_j$ does not depend on $i$ for $j\in S$. Therefore, the $(D,e)$-retrieval scheme also protects against the collusion pattern $\T=\binom{n}{\leq t} \cup \PowSet(I)$.
	%Thus, any collusion outside the information set $\supp(e)$ can be dealt with, as the corresponding projections are all from an element in $D$. 
	We use this insight to describe an optimal information set to use $t$-PIR for arbitrary collusion.
	
	Let $[n]$ be the set of servers and  $\T$ be the collection of collusion sets. Let $I$ be an arbitrary  information set such that  
	\begin{equation} |I| \leq n-k-t+1, \label{InfoSet} \end{equation}
	where $t:=\max\{ |T| : T \in \T \text{ and } T \cap I \neq \emptyset \}$. 
	As in Example~\ref{Example1}, a $t$-PIR schemes that uses such an $I$ as an information set can protect against the collusions in $\T$. If the inequality \eqref{InfoSet} is not sharp, we can furthermore puncture both $C$ and $D$ in $n-k-t+1-|I|$ points outside the information set. This reduces the number $n$ of servers, while all other parameters are preserved, and hence the rate is increased.
	This gives us the following result.
	\begin{lemma}\label{lm:infoset}
		For a given collusion pattern $\T$ we can download the symbols from $I$ at a rate of $r(I,t)=\frac{|I|}{|I|+k+t-1}$, where $I$ and $t$ are as in \eqref{InfoSet}.
	\end{lemma} 
	
	The rate in Lemma~\ref{lm:infoset} is clearly increasing in $|I|$ when $t$ is fixed. For fixed $t$, we can choose $I=I_t$ of maximal size by assigning \begin{equation}\label{eq:smallcollude}
	\tilde{I}_t\stackrel{\rm{def}}{=}[n]\setminus\bigcup_{T\in \T, |T|>t} T.\end{equation} We then let $I_t=\tilde{I}_t$ if $|\tilde{I}_t|\leq n-k-t+1$, and otherwise we let $I_t$ be an arbitrary $(n-k-t+1)$-element subset of $\tilde{I}_t$.
	
		To exploit Lemma~\ref{lm:infoset} to download an entire file, we need there to be a subset $S\subseteq[n]$ on which $C$ has full rank, such that $S$ does not intersect any colluding sets of size $>t$. In other words, the code $C$ has to have full rank on $\tilde{I}_t$. In fact, we can download an entire file at the rate of Lemma~\ref{lm:infoset} if and only if $C$ has full rank on $\tilde{I}_t$. Indeed, assume that $C$ has full rank on $\tilde{I}_t$, and that the file is subdivided into $r\stackrel{\rm{def}}{=}|\tilde{I}_t|$ blocks, each of which is encoded via $C$. Then we can repeat the $(D,e)$ download scheme $k$ times, each time downloading one new coded symbol from each block, such that all the downloaded symbols are from $\tilde{I}_t$. Hence, we use the same code $D$ but different vectors $e$ in each round.  Details on how to do this in the more restrictive setting of $t$-PIR are given in~\cite{razan_salim}. Note that, if the storage code $C$ is MDS, then the criterion that $C$ has full rank on $\tilde{I}_t$ is equivalent to $|\tilde{I}_t|\geq k$.
		
		From the scheme described above, together with Lemma~\ref{lm:infoset}, the following theorem follows:
		
		\begin{theorem}\label{thm:infoset}
		Let $C$ be an MDS storage code, and let $\T$ be a collusion pattern. Let $t$ be a positive integer such that $|\tilde{I}_t|\geq k$, where $\tilde{I}_t$ is as defined in \eqref{eq:smallcollude}. Then, we can download files privately from $C$ at rate \[\min\left\{\frac{|\tilde{I}_t|}{|\tilde{I}_t|+k+t-1} , \frac{n-k-t+1}{n}\right\}.\]
	\end{theorem} 
	\begin{IEEEproof}
	 Let $I\subseteq\tilde{I}_t$ be such that $|I|\leq n-k-t+1$. By Lemma~\ref{lm:infoset}, we can download symbols from $I$ privately at a rate $r(I,t)=\frac{|I|}{|I|+k+t-1}$. From the scheme described preceding this theorem, we can then download entire files at this rate, as $k\leq |\tilde{I}_t|$. If we select $I={I}_t$, we get $$\frac{|I|}{|I|+k+t-1}=\min\left\{\frac{|\tilde{I}_t|}{|\tilde{I}_t|+k+t-1} , \frac{n-k-t+1}{n}\right\}$$ by construction. This proves the theorem. 
	\end{IEEEproof}
	
It remains to select the optimal value of $t$. To this end, observe that $\tilde{I}_t$ is an increasing set in $t$. In particular, there exists a largest value $t_c$ of $t$ such that $|\tilde{I}_t|\leq n-k-t+1$. Observe that when $t\geq t_c$\,, then $$r(I_t,t)=\frac{n-k-t+1}{n}$$ is decreasing in $t$, and when $t\geq t_c$\,, then $$r(I_t,t)=\frac{\tilde{I}_t}{\tilde{I}_t +k+t-1}$$ is increasing in $t$. As a consequence, maximizing the rate $r(I_t)$ amounts to finding $t_c$, which we do by successively decreasing $t$ from $t_0=\max\{|T|:T\in\T\}$.
	
	\begin{example}
	Consider a $[6,2]$-storage code and let $\T=\langle\{1, 2\},\{3,4,5,6\}\rangle$. Then $\tilde{I}_2=\tilde{I}_3=\{1,2\}$ and $\tilde{I}_4=[6]$. As $|\tilde{I}_3|=2 \leq n-k-t+1$ and $|\tilde{I}_4|=6>n-k-t+1$, we get $t_c=2$. Therefore, the rate obtained by protecting against $4$-collusion is not optimal, but can be improved by using a $2$-PIR scheme, where the information set is selected to be $I_2=\{1,2\}$. The PIR scheme can now be applied to any subset of the servers of size $I_t+k+t-1=5$, as long as it contains $I_2$. Choosing not to use the $6^{th}$ node at all, we can achieve a rate of $r(I_2,2)=\frac{2}{2+2+2-1}=\frac{2}{5}$.
	% The resulting PIR rate is $\frac{2}{6}$. A further gain can be achieved by considering a punctured version of the storage code, \emph{e.g.}, by not using the $6^{th}$ node at all. We then are in a situation analogous to \Cref{Example1} and can achieve a rate of $r(I_2,2)=\frac{2}{2+2+2-1}=\frac{2}{5}$.
	\end{example}
	
	\section{Partitions of Colluding Sets}
Next, we will discuss the case where the maximal elements of $\T$ are a partition of the set $[n]$ of servers. This means that there is a collection $T_1,\cdots, T_r$ of colluding sets, such that no collusion is happening between $T_i$ and $T_j$ when $i\neq j$. We begin with an example suggesting how to improve the scheme from Section~\ref{sec:partial}.
	\begin{example}
		Let $C$ be an MDS storage code on $n=6$ nodes, and $\T$ be generated by the sets $\{1, 2, 3\}, \{4, 5, 6\}$. Using a $3$-PIR scheme we could achieve a rate of $\frac{n-t-k+1}{6}=\frac{4-k}{6}$. 
		For $k=3$, using a $3$-PIR scheme would thus allow the user to download $1$ part per subquery, achieving a rate of $\frac{1}{6}$.
		On the other hand, using a scheme of the form $(\rep,e)$, the user would be able to download the whole file, i.e. $3$ parts in one query. A sketch of the scheme applied is shown below.
		
		Suppose the files are stored using an $(n,k)$-MDS code, and the user wants file $x^f$.
		The user can then send a random vector $u$ to one set of colluding nodes, say to nodes $1, 2,$ and $3$, and send $u+e_f$ to the nodes $4,5,$ and $6$.
		
		The user will then decode the randomness, i.e. the projection of the random vectors, from the responses of nodes $1,2,3$. Then, substituting these in the responses from nodes $4,5,$ and $6$, will give the user $3$ independent equations in $x^f(1), x^f(2),$ and $x^f(3)$, thus allowing her to decode the file $x^f$.
		
		This achieves rate $\frac{3}{6}=\frac{1}{2}$.
	\end{example}
	
	In general, consider the case where $D=\rep_n$ is the repetition code, as in Corollary~\ref{thm:disconn}. In this case, the query vector $e$ must be chosen to be constant on each colluding set $T\in \T$. The $(D,e)$-retrieval scheme to obtain file $x^i$ will then retrieve a uniformly random vector from $C\star\rep+ x^i(C\star e)$. After decoding this with a generator matrix for $(C\star\rep)^\perp=C^\perp$, the decoded vector will be \begin{equation}\label{eq:retrieved} x^i(C\star e)C^\perp=x^i(C\diag(e)C^\perp). \end{equation} Here, $\diag(e)$ is the $n\times n$ diagonal matrix with the vector $e$ on the diagonal, and by abuse of notation, we use $C$ and $C^\perp$ to denote some chosen generator matrix and parity check matrix for the code $C$, respectively. From~\eqref{eq:retrieved}, it follows that the number of downloaded symbols is
\vspace{-0.5em}	 \[\rank(C\diag(e)C^\perp)\leq%\min\{\rank(C),\rank(\diag(e)),\rank(C^\perp)\}=
	 \min\{k,w(e),n-k\},\] 
	 where $w(e)=|\supp(e)|$ is the weight of the vector $e$. It also follows that we may without loss of generality assume that $e$ is a $0$-$1$-vector, since multiplying $\diag(e)$ with a non-degenerate diagonal matrix does not change the rate of the composed matrix.
	
After possibly reordering the coordinates, we may assume that $e=[1, \cdots , 1, 0, \cdots , 0]$ has ones in the first $w=w(e)$ coordinates, and zeroes in the last $n-w(e)$ coordinates. If $C$ is an MDS storage code, we may also assume that the generator matrices are chosen such that \[C=[ I , M] \hbox{ and } C^\perp=\underbrace{\begin{bmatrix}M^T \\ -I\end{bmatrix}}_{n-k}.\] Assuming $w(e)\geq k$, we then get \[\diag(e)\cdot C^\perp=
\left[\begin{array}{cc} \multicolumn{2}{c}{\quad M^T} \\  -I_{w(e)-k}& 0\\
0 &0 
\end{array}\right],\] with zeros in the $n-w(e)$ last rows. The first $w(e)-k$ columns of $\diag(e)\cdot C^\perp$ agree with those of $C^\perp$, so will be annihilated when left-multiplied with $C$. It follows that the rank of $(C\star e)C^\perp$ is not more than $n-w(e)$. On the other hand, the last $n-w(e)$ columns in $\diag(e)\cdot C^\perp$ are not orthogonal to $C$, so we have $\rank((C\star e)C^\perp)= n-w(e)$ if $n-w(e)\leq k\leq w(e)$ We have thus proven the following.
		\begin{theorem}
		Let $C$ be an MDS storage code. Then the $(\rep,e)$-retrieval scheme has rate \[\frac{\min\{k,n-k, w(e), n-w(e))\}}{n}.\]
		\end{theorem}
		Applying this scheme in the case where the collusion pattern is a partition, we want to maximize $\min\{w(e), n-w(e)\}$, subject to $e$ being constant on each collusion set. In other words, $\supp(e)$ has to be a union of collusion sets, of size as close to $n/2$ as possible. The following result is immediate, and describes the important case where the coding rate of $C$ is at least $1/2$, or in other words $k\geq n-k$:
		\begin{theorem}\label{thm:partitionrate}
		Let $\T\subseteq\langle T_1, T_2\rangle$ be a collusion pattern where $T_1$ and $T_2$ are disjoint sets, each of size $\leq k$. Then there exists a  vector $e$ such that the $(\rep, e)$-retrieval pattern has retrieval rate $\frac{n-k}{n}$.
		\end{theorem}

	In the special cases where $k<n-k$, another subtlety occurs. In such cases, it is favorable to subdivide the files into so-called \emph{stripes}, following the ideas in~\cite{Extended}. This allows us to simultaneously download several stripes of the same file, each of which has size $k$, thereby circumventing the $k/n$ lower bound for the retrieval rate. The key elements of this is shown in the example below.
		\begin{example}
		Let $n=9$ and $k=3$ and $\T$ be generated by the sets $\{1, 2, 3\}, \{4, 5, 6\}, \{7, 8, 9\}$.
		Here, we notice that it would be helpful to split our files into $2$ stripes. %, i.e. we can look at each file $x^i$ as a $k\times 2$ matrix. % $\left[ \begin{array}{cc}
%		x^i(1,1) & x^i(1,2) \\
%		\dots & \dots \\
%		x^i(k,1) & x^i(k,2)
%		\end{array}\right].$ 
The entire data matrix is thus a $k\times 2m$ matrix $$X=\left[ \begin{array}{cccccc}
		x^1(1,1) & x^1(1,2) & \dots & x^m(1,1) & x^m(1,2)\\
		\dots & \dots \\
x^1(k,1) & x^1(k,2) & \dots & x^m(k,1) & x^m(k,2)
		\end{array}\right].$$
		
		Suppose the user wants file $x^f$.
		The user can then send a random vector $u$ to one set of colluding nodes, say to nodes $1, 2,$ and $3$, send $u+e_{2f-1}$ to the nodes $4,5,$ and $6$, and send $u+e_{2f}$ to the nodes $7,8,$ and $9$.
		
		The user will then decode the randomness, i.e. the projection of the random vectors, from the responses of nodes $1,2,3$. Then, substituting these in the responses from nodes $4,5,$ and $6$, will give the user $3$ independent equations in $x^f(1,1), x^f(2,1),$ and $x^f(3,1)$, thus allowing him/her to decode the first stripe of the file $x^f$. Doing the same with the responses from nodes $6,7,$ and $8$,  user can decode the second stripe of the file. Thus, the user will decode the file $x^f$ in one query.
		
		This achieves rate $\frac{6}{9}=\frac{n-k}{n}$, extending the result from Theorem~\ref{thm:partitionrate} to the case where $k<n-k$.
	\end{example}

	\section{Conclusion}

We have studied private information retrieval schemes for data encoded across several servers, wherein an arbitrary collection of subsets of the servers are colluding.  We applied the so-called \emph{$(D,e)$-retrieval scheme} from \cite{FGHK16}, which protects against the collusion of all subsets of a given size $t$, to the present case of arbitrary collusion pattern.  By carefully choosing the information set $I$ and the amount of collusion $t$ to protect against, we can offer an improvement in the PIR rate over the naive strategy of using a $(D,e)$-retrieval scheme to protect against the largest colluding set.   Finally, we investigated achievable PIR rates when the sets of colluding servers form a partition of the set of all servers.  Future work will consist of deriving precise expressions for the achievable rates for arbitrary collusion patterns.

	\bibliographystyle{IEEEtran}
	\bibliography{references,coding2,coding1}
\end{document}